\documentclass[review,11pt]{elsarticle}

\usepackage{color}
\usepackage{hyperref}
\usepackage{graphicx}

\begin{document}

\begin{frontmatter}

\title{Design and characterization of a W-band system for modulated DNP experiments}

\author{Mallory L. Guy}
\author{Lihuang Zhu}
\author{Chandrasekhar Ramanathan\corref{cor1}}
\ead{chandrasekhar.ramanathan@dartmouth.edu}

\address{Department of Physics and Astronomy, Dartmouth College, Hanover NH 03755, USA}

\cortext[cor1]{Corresponding author}

\begin{abstract}
Magnetic-field and microwave-frequency modulated DNP experiments have been shown to yield improved enhancements over conventional DNP techniques, and even to shorten polarization build-up times. The resulting increase in signal-to-noise ratios can lead to significantly shorter acquisition times in signal-limited multi-dimensional NMR experiments and pave the way to the study of even smaller sample volumes.  
In this paper we describe the design and performance of a broadband system for microwave frequency- and amplitude-modulated DNP that has been engineered to minimize both microwave and thermal losses during operation at liquid helium temperatures. The system incorporates a flexible source that can generate arbitrary waveforms at 94 GHz with a bandwidth greater than 1 GHz, as well as a probe that efficiently transmits the millimeter waves from room temperature outside the magnet to a cryogenic environment inside the magnet.  Using a thin-walled brass tube as an overmoded waveguide to transmit a hybrid HE11 mode, it is possible to limit the losses to 1 dB across a 2 GHz bandwidth.  The loss is dominated by the presence of a quartz window used to isolate the waveguide pipe.  This performance is comparable to systems with corrugated waveguide or quasi-optical components.
The overall excitation bandwidth of the probe is seen to be primarily determined by the final antenna or resonator used to excite the sample and its coupling to the NMR RF coil.  Understanding the instrumental limitations imposed on any modulation scheme is key to understanding the observed DNP results and potentially identifying the underlying mechanisms.  We demonstrate the utility of our design with a set of triangular frequency-modulated DNP experiments.
\end{abstract}

\end{frontmatter}


\section{Introduction}

Dynamic nuclear polarization (DNP) offers one route to overcoming the low detection sensitivity of NMR that has frequently limited its applicability, requiring either large sample volumes or long data acquisition times to attain acceptable signal-to-noise ratios.  Microwave-induced DNP has been applied to solid-state biomolecular spectroscopy at cryogenic temperatures \cite{Hall-1997,vanderWel-2006,Maly-2008,Mak-Jurkauskas-2008,Bajaj-2009,Salnikov-2010,Sergeyev-2011,Jasco-2012,Potapov-2013}.  Dissolution DNP techniques \cite{Ardenkjaer-Larsen-2003} have enabled hyperpolarized solutions to be used for room-temperature studies in both process engineering \cite{McCarney-2007} and biomedical applications \cite{Golman-2006,Day-2007,Gallagher-2008,Gallagher-2009,Kurhanewicz-2011,Cassidy-2013b,Nelson-2013}.  The large spin polarizations have also enabled high temporal resolution studies of chemical processes \cite{Bowen-2008}, the use of novel ultra-fast multidimensional spectroscopy techniques \cite{Frydman-2007,Panek-2010}, and could possibly allow extremely high-resolution magnetic resonance imaging \cite{Thurber-2010}.

The enhancements achieved in typical continuous wave (CW) DNP experiment are often as much as an order of magnitude lower than the theoretically predicted value of the ratio of the electron gyromagnetic ratio to the nuclear gyromagnetic ratio \cite{Abragam-1978}. 
There are many factors that limit the enhancements observed in CW DNP experiments, such as imperfect microwave saturation caused by limited microwave power, inefficient excitation of a broad ESR line, and the presence of multiple DNP and leakage pathways. A number of efforts have been made to improve DNP enhancements by addressing each of these issues.  Of particular note are the efforts to develop water soluble free radicals that optimize DNP enhancement at high magnetic fields \cite{Macholl,Song,Matsuki,Sauvee-2013}.  Magnetic field modulation \cite{Tycko} and microwave frequency modulation \cite{Cassidy-2013b,Hovav,Bornet} have both been used to improve DNP enhancements by a factor of 2--3  under non-spinning conditions.    Bornet et al.\ have observed a dramatic reduction in DNP build-up times with frequency modulation \cite{Bornet}.  While one can naively imagine that the frequency modulation helps saturate an inhomogenously-broadened EPR line, the detailed physical mechanisms for the enhancement are still being elucidated \cite{Hovav}.

In MAS-DNP experiments, DNP enhancements have been shown to depend on rotor-speed \cite{Rosay-2010}, potentially arising from adiabatic sweeps and induced level crossings in the rotating frame \cite{Thurber-2012,Mentink-Vignier-2012}.  As the underlying mechanisms behind both MAS-DNP and frequency-modulated DNP are elucidated, it is conceivable that we could create rotor-synchronized modulated DNP excitation schemes.

While the presence of leakage relaxation pathways has long been known \cite{Abragam-1978}, the possibility of multiple (potentially competing) DNP pathways has recently also been pointed out \cite{Banerjee-2013,Shimon-2014,Can-2014}.
Cory and co-workers have explored the use of  optimal control techniques to improve pulsed DNP enhancements at low field  by selectively exciting a particular DNP pathway while suppressing leakage paths \cite{Sheldon}.  These studies suggest that amplitude and frequency modulated microwave excitation schemes hold potential to improve the enhancements obtained in DNP experiments.  However, these techniques also impose additional design constraints on the microwave transmission schemes used in DNP probe design.

In this paper we describe the design and performance of a W-band DNP system, that can be used for both pulsed and CW DNP and can readily implement both frequency and amplitude modulated DNP experiments.  The system consists of a programmable millimeter wave source positioned outside the magnet, as well as a cryogenic probe that is designed to fit inside a Janis STVP continuous flow cryostat, which in turn fits in the bore of a 3.34~T ($\approx$ 94 GHz ESR frequency for $g \sim 2$) Oxford superconducting NMR magnet (100 mm superwide bore). In this system the magnet center is located 735 mm below the top flange of the magnet, and the top of the cryostat is 400 mm above the top flange of the magnet.

\section{Millimeter-wave subsystem}
\subsection{Frequency Source}
\noindent 
Two schemes have typically been used to implement frequency modulation in DNP experiments; either modulating the input to a varactor-tuned Gunn diode \cite{Cassidy-2013b,Hovav} or modulating the input to a frequency-multiplier-based source such as the ELVA-1 system \cite{Bornet}.  Our souce, which is similar in many respects to the systems implemented at the Weizmann Institute \cite{Gromov-1999,Feintuch-2011}, is specifically designed to engineer arbitrary waveform shaping capabilities at millimeter wave frequencies for both pulsed and CW excitation.

\begin{figure}
  \centering
    \includegraphics[width=1\textwidth]{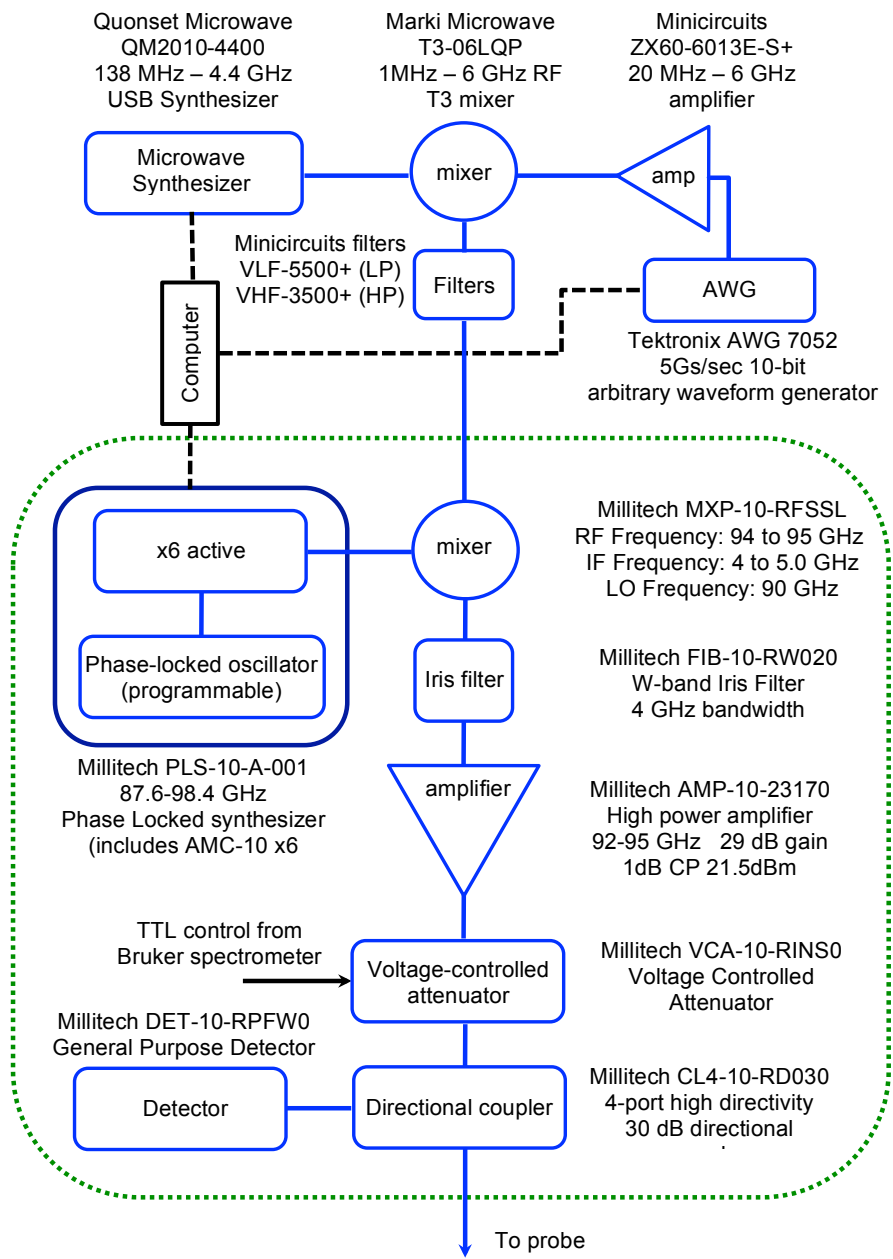}
  \caption{Block diagram of microwave source used in the frequency modulated DNP experiments. A detailed discussion of the system is provided in the text.}
    \label{fig:schematicofsource}
\end{figure}

A schematic of the source is shown in Figure~\ref{fig:schematicofsource}.  It consists of a programmable phase-locked millimeter wave synthesizer (Millitech PLS-10-A-001) that can output frequencies in the range of 87.6 to 98.4 GHz.  The synthesizer combines a lower-frequency programmable phase-locked oscillator with an $\times6$ active multiplier.  This signal, which is set to about 90 GHz, is fed into the local oscillator port of a balanced mixer (Millitech MXP-10-RFSSL).  The intermediate frequency (IF) port on the mixer has a bandwidth of 4-5 GHz, which is where we input a modulated 4 GHz signal as described below.  The output of the mixer is passed through an iris filter with a 4 GHz bandwidth that filters out the lower side band at 86 GHz (Millitech FIB-10-RW020), and is passed to a 200 mW high power amplifier (Millitech AMP-10-23170) .  The signal is then passed through a voltage controlled variable attenuator (Millitech VCA-10-RINS0) that is used for gating the millimeter wave pulses, but could potentially also be used for additional modulation.  We currently use a programmable TTL output line from the Bruker spectrometer (with an appropriate level shifter to match the 10 V range of the attenuator) to attenuate the microwaves and minimize noise pickup during acquisition of the NMR FID.  The millimeter waves then pass through a 30 dB directional coupler, which is used to monitor the microwave excitation, before being transmitted to the probe.

In order to generate the modulated 4 GHz waveforms, we mix the output of a constant 3.5 GHz signal generated by a programmable USB synthesizer (Quonset Microwave QM2010-4400) with the (amplified) signal generated by a high-frequency 
arbitrary waveform generator (Tektronix AWG-7052).  The signal is then filtered and input to the IF port of the millimeter wave mixer.  For larger frequency modulations, we lower the frequency of the USB synthesizer and adjust the AWG frequency accordingly so that all modulations are in the 4 GHz window.  A computer is used control the frequency of the USB synthesizer, trigger the AWG, and program the frequency of the millimeter wave synthesizer.  Note that this system can readily be adapted to build a versatile pulsed EPR spectrometer with the addition of a receiver stage.

\subsection{Transmission}
\subsubsection{Overview}
\noindent One of the biggest challenges in designing a DNP probe is transmitting high frequency microwaves efficiently over long distances while simultaneously minimizing thermal losses. While the solid-state sources discussed above offer significant advantages in terms of programmability and frequency-agility, they are fundamentally limited in terms of the available power that they are able to deliver.  It is therefore critical to minimize the millimeter wave losses during transmission into the cryostat.
There are many approaches to this, including the use of corrugated waveguides, standard over-moded waveguides, quasi-optical designs, and fundamental mode waveguides \cite{Tycko,Cho,Corrugated,Woskov,Barnes,Songi,Leggett-2010,Pike}.  

While thermal engineering is less critical for DNP experiments performed at liquid nitrogen temperatures, it is very important in the design of experiments at liquid helium temperatures.  
In general, fundamental mode waveguide is the simplest scheme to set up, but there is a tradeoff between minimizing microwave and thermal losses (see Table 1).  For example, copper waveguides minimize microwave losses but have the highest thermal losses, while stainless steel waveguides have high microwave losses but low thermal losses.
Since the millimeter waves only propagate through a thin surface layer of the waveguide, gold or silver plating of a low-thermal conductivity metal such as stainless steel would seem a reasonable solution.  However, it is typically difficult to gold- or silver-plate such narrow structures uniformly over the approximately one meter long length required, unless it can be performed in sections that are then aligned carefully.

The use of over-moded waveguide improves the microwave performance over fundamental mode.  Since conventional over-moded waveguides are larger than fundamental mode waveguides, they have a higher thermal mass and thus present more of a problem with heat loss during experiments at liquid helium temperatures.  Additionally, they may not be compatible with certain experimental set-ups because of space constraints.  
Corrugated waveguides and quasi-optical approaches to microwave transmission have shown the lowest losses to date. Quasi-optical setups can suffer from stability issues since the components are sensitive to temperature \cite{Nanni}. Decreases in microwave power delivered to DNP samples have been reported to be as high as 50$\%$ with an angular misalignment of just 0.1$^\circ$ \cite{HanPCCP}.  Table 1 summarizes the loss experienced by each of these common types of transmission at W-band.	

\begin{table}
\small
	\begin{center}
    \begin{tabular}{ | l | c| c | p{2cm} |}
    \hline
    Transmission scheme & W-band microwave loss  & Thermal conductivity \\
    & (dB/m) & (W/m K) \\ \hline
    Fundamental (copper) & 2.59 & 380 - 390 \\ \hline
    Fundamental (silver coated) & 2.59 & 406 - 418\\ \hline
    Fundamental (stainless steel) & 18.86 & 16 - 18\\ \hline
    Over-moded (copper) &  1-1.5 &  380 - 390\\ \hline
    Corrugated (brass) & $<$ 1 & 109 - 121\\ \hline
    Quasi-optical designs  & $<$ 1 & \\ 
    \hline
    \end{tabular}
    \caption{Summary of the loss and thermal conductivity of common methods of W-band  microwave transmission \cite{Nanni,Moreno}.}
\end{center}
\end{table}

\subsubsection{Design}
\begin{figure}
  \centering
    \includegraphics[trim = 25mm 120mm 55mm 20mm, clip,width=.75\textwidth]{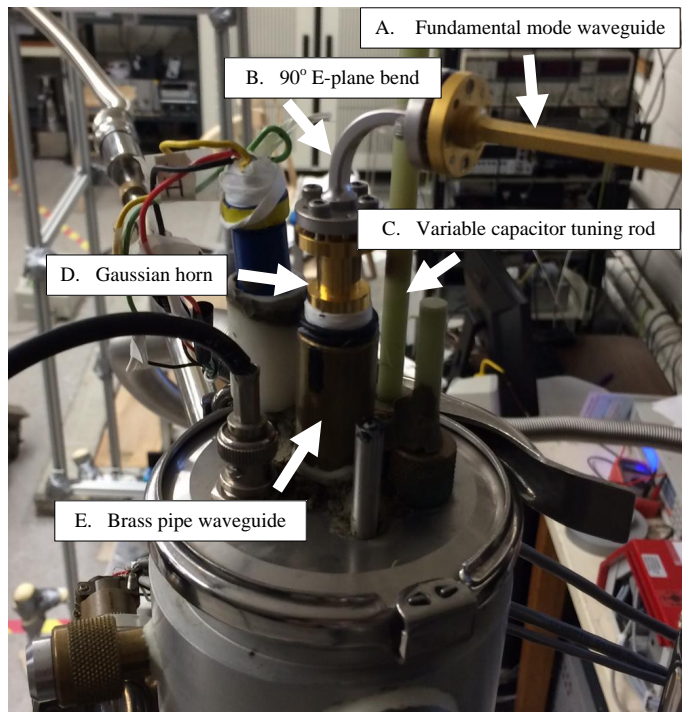}
  \caption{Top of our DNP microwave transmission, showing the 24-inch (610 mm) section of gold-plated fundamental mode waveguide which extends over the center of the magnet (A); the 90 degree E-plane bend which orients the waveguide along the axis of the magnet (B); the tuning rod for the variable capacitor used to tune the RF NMR coil (C); the scalar horn antenna used to convert from fundamental mode to a hybrid HE11 mode (D); and the brass pipe waveguide (E).} 
  \label{fig:DNPprobephoto}
\end{figure}

\noindent Our experimental set-up is a relatively simple, cost-effective approach to balancing the competing requirements of  microwave transmission and thermal efficiency.   The top of the microwave  transmission setup is shown in Figure~\ref{fig:DNPprobephoto}.  The source is connected to a 
 24 inch (610 mm) section of gold-plated fundamental mode waveguide which extends over the center of the magnet (A).  This is followed by a 90 degree E-plane bend which orients the waveguide along the axis of the magnet (B).  After the 90 degree bend, the fundamental mode is converted to a hybrid HE11 mode using a scalar horn antenna (SFH-10-R0000 Millitech Inc.).  

The output of the horn sits inside a thin-walled brass pipe (5/8-inch or 15.875 mm diameter, 0.05-inch or 1.27 mm wall thickness) which acts as an over-moded waveguide and extends down the length of the probe.  The top of the brass pipe is sealed by a  1 mm-thick quartz window using an epoxy seal to isolate the interior of the cryostat during cryogenic experiments.
The  hybrid HE11 mode has an exponentially decaying profile that suppresses the fields (and currents) at the wall of the waveguide, minimizing microwave losses. The thin-walled brass pipe also has a small thermal mass that limits heat losses.  We estimate the thermal loss in our system to be similar to that of the corrugated brass waveguide in Table 1.
Note that this use of a brass pipe as an overmoded waveguide for a hybrid HE11 mode is distinct from the use of circular waveguide carrying a fundamental TM11 mode or an overmoded TE11 mode \cite{Bornet,Comment-2007,Granwehr}.    Groups using corrugated waveguide to transmit millimeter waves have informally reported using sections of cylindrical pipe as waveguide sections in their setups with little loss in performance.

\begin{figure}
  \centering
    \includegraphics[trim = 13mm 155mm 55mm 32mm, clip,width=1\textwidth]{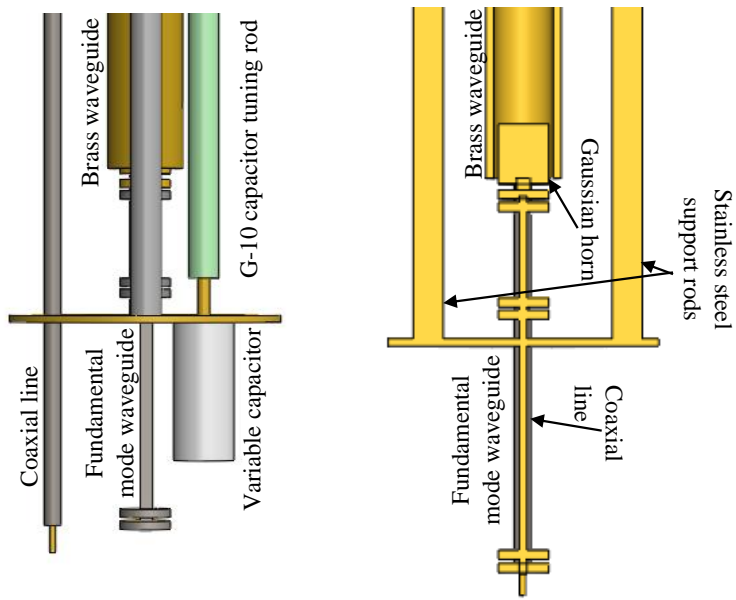}
  \caption{Side views of the bottom of the probe, showing the relative positions of the RF coaxial cable, the brass waveguide and the receiving scalar horn antenna, and the variable capacitor and tuning rod.}
 \label{fig:DNPprobeschematic}
\end{figure}
 
About 6 inches (152 mm) from the center of the magnetic field, the brass pipe ends and a receiving scalar horn antenna sits inside the brass pipe. This horn re-converts the microwaves back to fundamental mode.  The conversion back to fundamental mode is done to achieve better control of the orientation of the microwave magnetic field at the sample, as will be discussed in the next section.  The receiving horn is followed by a short section ($\approx$ 4 inches or 101 mm) of fundamental mode waveguide that extends to just above the NMR coil, as shown in Figure~\ref{fig:DNPprobeschematic}.  At this waveguide flange, it is possible to connect either a broad-band antenna such a standard pyramidal scalar gain antenna or a resonant structure consistent with DNP measurements such as a Fabry Perot cavity \cite{Singel-1989,FabryPerot} or a scroll resonator \cite{Weis-1999}.

\begin{figure}
  \centering
    \includegraphics[trim =10mm 10mm 10mm 28mm, clip,width=.95\textwidth]{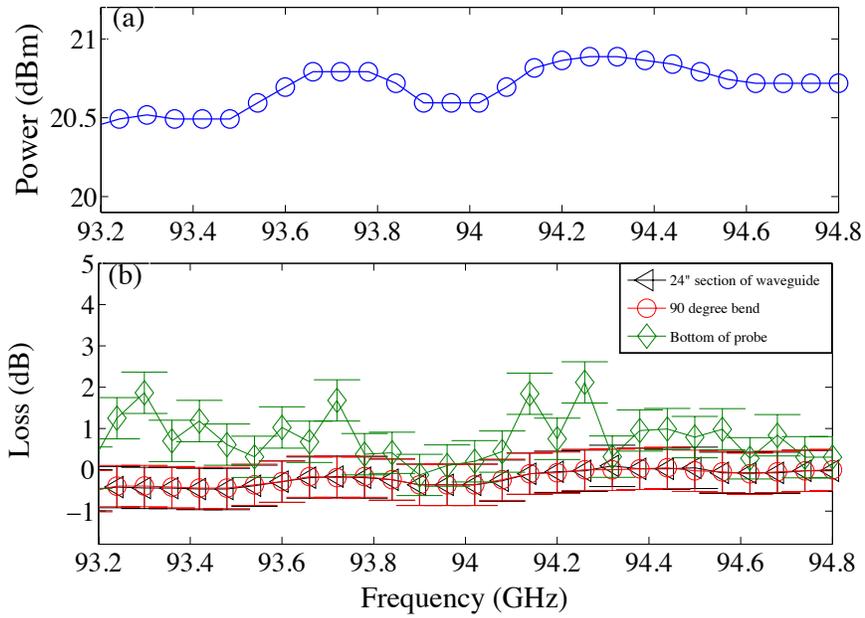}
  \caption{  
  (a) Output power of the microwave source as a function of frequency. The measured output power of the source is 112 mW (20.5 dBm).
  (b) Loss as a function of frequency for each stage of transmission down the DNP probe. The black triangles represent the loss after the 24'' section of waveguide, the red circles represent the loss after the 90$^\circ$ bend, and the green diamonds represent the cumulative loss at the bottom of the probe. The apparent negative loss values, are an artifact of experimental uncertainties in our measurement.  All the apparently negative values are within a single standard deviation (approximately 0.4 dB) of lossless transmission.}
  \label{fig:PowerPlotPaper}
  \end{figure}

\subsubsection{Performance}
\noindent 
We measured the insertion loss at each stage of transmission using a W-band diode detector (Spacek DW-2P) in the 92--96 GHz frequency range.  This range was motivated by our goal to explore millimeter-wave frequency modulations in the range of $\pm1.5$ GHz about the 94 GHz center frequency --- a range that is comparable to the width of the broadest ESR linewidths and hyperfine interactions we expect to explore.

First, the detector was connected directly to the output of the microwave source to record the microwave power output by the source as a function of frequency, as shown in Figure~\ref{fig:PowerPlotPaper}(a).  The detector was then used to measure the microwave power as a function of frequency at three locations along the probe:  at the end of the long section of the fundamental mode waveguide; after the 90 degree E-plane bend; and at the end of the receiving scalar horn antenna at the bottom of the probe.  Figure~\ref{fig:PowerPlotPaper}(b) shows the cumulative losses measured up to each of these locations.  Losses along the fundamental mode waveguide and the 90 degree bend were found to be negligible in the frequency range  measured.  The apparent negative loss values, are an artifact of experimental uncertainties in our measurement.  All the apparently negative values are within a single standard deviation (approximately 0.4 dB) of lossless transmission.  The loss measurements of the brass overmoded waveguide were taken with the probe positioned vertically on a probe stand.   In this configuration small adjustments and movements of the probe were observed to cause fluctuations in the measured power (indicated by the relatively large error bars).  Small movements can cause the scalar horn antennas to shift relative to each other, and even a slight misalignment in the transmitting and receiving horns can result in significant power loss.  We note that this instability and fluctuation should be minimized when the probe is inside the magnet during experiments. A future improvement to this transmission scheme would be the development of a better method of aligning the two horns.

The loss of the entire transmission scheme is typically less than 1 dB, including the mode-conversions and the long section of fundamental mode waveguide.  It is important to note that this loss is dominated by the quartz window.  The thickness of sapphire and quartz windows has previously been shown to significantly influence transmission at millimeter wave frequencies \cite{Jawla-2012}.  A better optimized window (improved thickness, material, angle) would reduce losses even further. When we repeated the measurements using a brass tube that did not have a window, the transmission was observed to be almost lossless.  The loss profile is approximately frequency independent over the range 93-95 GHz.  The performance of this system is thus comparable to corrugated waveguide and quasi-optical schemes.

\subsection{Microwave Sample Excitation}
\noindent The antenna or cavity structure used to irradiate the sample depends on the coil and sample geometry used.  
Since only the microwave magnetic field components transverse to the main static field contribute to the excitation of the electron spins, it is important that the microwave power delivered down the probe is efficiently coupled to a mode that has such a transverse component.  

We used Ansoft HFSS to perform three-dimensional finite-element simulations of the distribution of the electromagnetic fields to examine the effect of the RF NMR coil when either the transmitting waveguide pipe, or a simple pyramidal scalar gain horn is used to excite the electron spins in the sample. 

For the simulations with the pyramidal horn, we specify the dimensions of the exciting port, which has the dimensions of fundamental mode WR-10 waveguide, and the power that is input at the port (1 Watt).  For the simulations with the cylindrical pipe, we specify that a Gaussian mode is incident on the exciting port at the same 1 Watt power level.  The simulations are free space simulations with no external boundaries used, but we do specify a volume over which the calculations are performed.  The cylindrical pipe simulations are run within a cylindrical volume of $26.28$ cm$^3$. The volume of the free space region in the pyramidal horn simulations is $24.92$ cm$^3$.  All simulation results are shown on the same color scale, so that intensities can be easily compared across different images. The fixed frequency simulations were performed at 93.75 GHz.

\begin{figure}
  \centering
    \includegraphics[width=\textwidth]{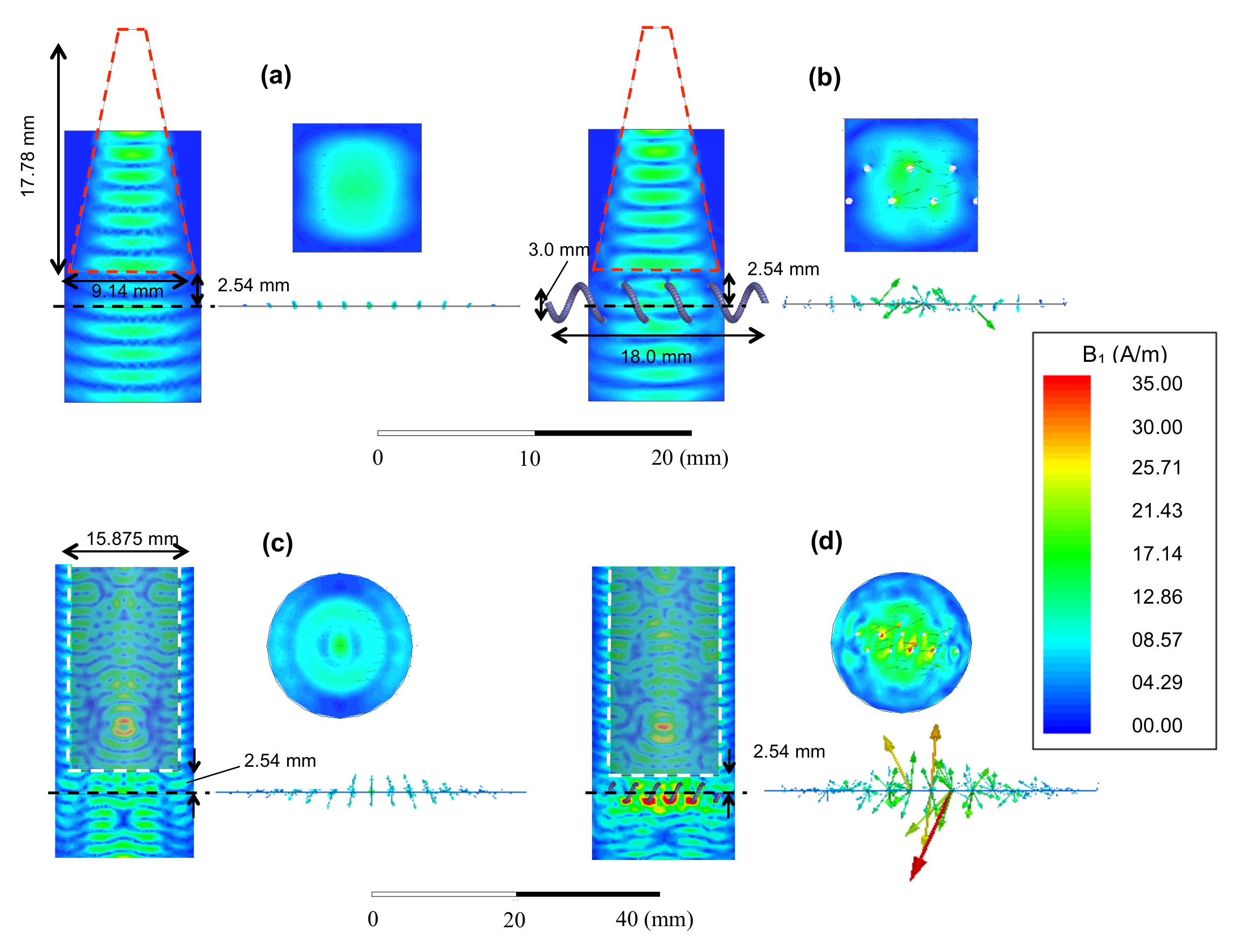}
  \caption{HFSS simulations of the magnetic field distribution for a pyramidal horn (a) without an RF coil and (b) with an RF coil, and for the cylindrical brass pipe (c) without an RF coil and (d) with an RF coil.   The figure also shows the cross-sectional intensity and orientation of the magnetic field at the sample plane (indicated by a dashed line) in each case. The simulations were performed at a frequency of 93.75 GHz with a nominal input power of 1 W.}
  \label{fig:HFSShornandpipe}
\end{figure}

Figure~\ref{fig:HFSShornandpipe}(a) and (b) show the magnetic field distributions for the pyramidal horn without and with the NMR coil.  A fundamental TE11 mode is seen to propagate at the top of the horn and gradually broaden as the wavefront leaves the horn. 
The 3 mm diameter coil is placed 0.1 inches (2.54 mm) outside the horn.  In each case, the images on the right show a spatial map of the magnetic field magnitudes and the orientation of the field at the sample position, which is indicated by the dashed black line.  Figure~\ref{fig:HFSShornandpipe}(c) and (d) show the corresponding field distributions when the cylindrical brass tube used to transmit the millimeter waves down the cryostat is also used to excite the sample.   Note that the scale bars sizes are different for the horn and pipe simulations, as the dimensions of the horn are significantly smaller than the pipe.    In the absence of the coil the pyramidal horn is seen to provide both better homogeneity, as well as control over the orientation of the magnetic fields.    In both cases the RF coil is seen to significantly alter the mode structure of the fields at the sample location.

\begin{figure}
\centering
\includegraphics[width=1\textwidth]{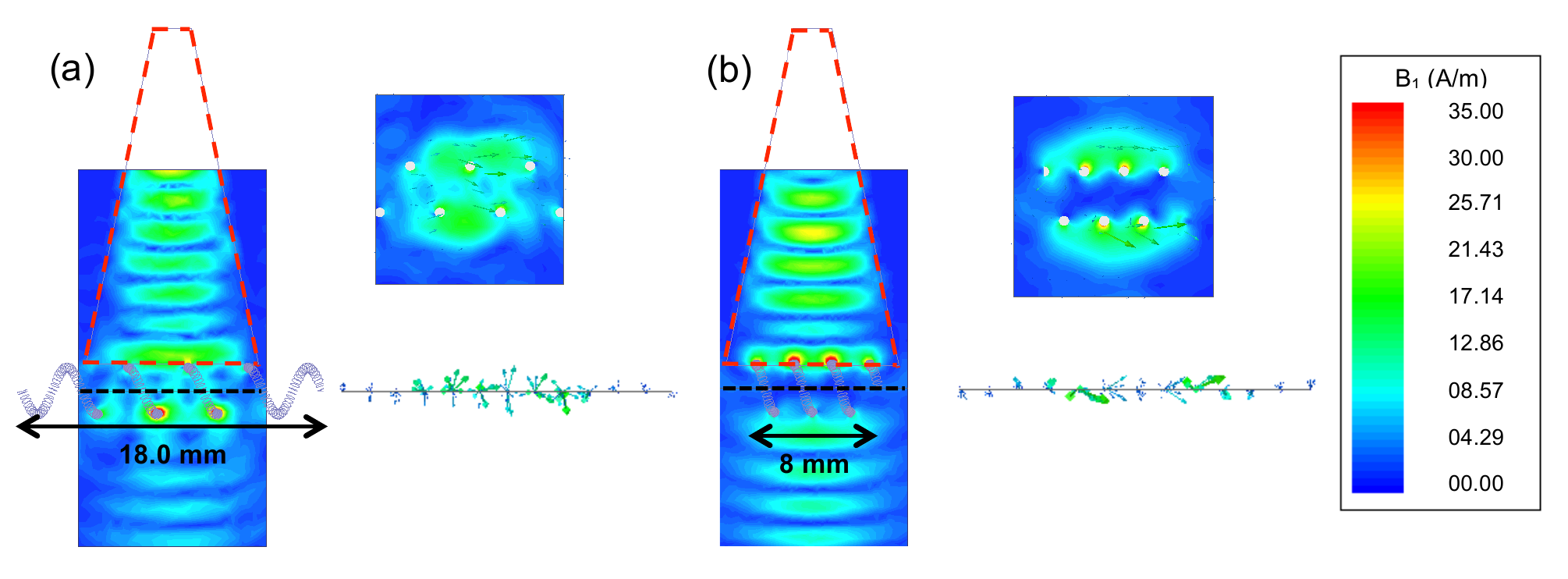}
\caption{HFSS simulations of the magnetic field distribution for a pyramidal horn and two RF NMR coils with slightly different lengths and pitches.  The figures also show (right) the cross-sectional intensity and orientation of the magnetic field at the sample plane (indicated by a dashed line) in each case.  The simulations were performed at a frequency of 93.75 GHz with a nominal input power of 1 W. }
\label{fig:HFSSScreening}
\end{figure}

The RF NMR coil can also lead to a screening of the millimeter waves out of the sample.  Figure~\ref{fig:HFSSScreening} shows the field distributions obtained using 
two RF coils with different lengths and pitches, where the diameter of the coils and their position relative to the horn remains fixed.  It can be seen that for the tighter winding the magnitude of the microwave magnetic field becomes negligible inside the coil. Such screening effects have been noted for low-frequency pick-up coils used in LOD EPR experiments \cite{Granwehr}.

\begin{figure}
\centering
\includegraphics[width=0.75\textwidth]{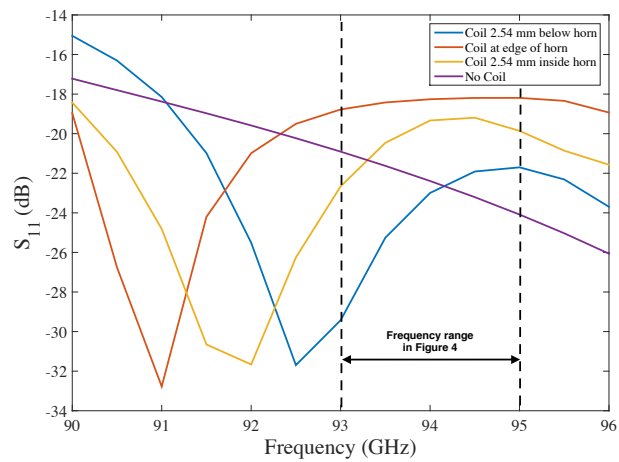}
\caption{HFSS simulations showing the simulated $S_{11}$  network parameter as a function of excitation frequency for the pyramidal horn without an NMR coil and for three different NMR coil positions relative to the horn. The $S_{11}$ network parameter measures the fraction of the power input into a network that is reflected back to the excitation port and provides information about the frequency response of the system. Note that the frequency range examined here is significantly wider than that characterized in Figure~\ref{fig:PowerPlotPaper}, which is indicated by the pair of dashed lines. 
}
\label{fig:HFSSFrequencyResp}
\end{figure}

Our probe uses a pyramidal horn for the millimeter wave excitation with the RF coil located just below the horn.   While both the brass cylindrical pipe and the pyramidal horn are broadband structures, the interaction with the RF coil can introduce a frequency dependence into their performance. Figure~\ref{fig:HFSSFrequencyResp} shows the variation in the simulated $S_{11}$ network parameter with frequency for the pyramidal horn without an NMR coil, and when the coil position is varied relative to the horn.   The $S_{11}$ network parameter measures the fraction of the power input into a network that is reflected back to the excitation port and provides information about the frequency response of the system.  It can be seen that the coupling between the coil and horn introduces a strong frequency dependent response.  Note that the frequency range examined here is significantly wider than that characterized in Figure~\ref{fig:PowerPlotPaper}, which is indicated by the pair of dashed lines.  For small frequency deviations this frequency dependence is not important, but it does become important for larger frequency modulations, or short pulses.
It should be noted that the Cory \cite{Borneman-2012} and Han \cite{Kaufmann-2013} groups have recently demonstrated that it is possible to compensate for bandwidth restrictions, and that these do not pose a fundamental limitation to the design of complex microwave modulation schemes.

\section{NMR coil}
\noindent In contrast to probes where the tuning elements are located remotely from the coil \cite{Conradi,Barnes-2009}, our system uses cryogenic tuning elements.  The NMR detection circuit used to collect the data is a standard series-match parallel-tune resonance circuit \cite{Nuts}. A semi-rigid coaxial cable with stainless-steel outer conductor and a silver plated copper inner conductor (Micro-coax -UT-141-SS) connects the room temperature BNC connector to the tuned low-temperature circuit.  The matching capacitor is a low temperature, fixed value capacitor (3 pF). The tunable capacitor is a ceramic hermetically sealed capacitor (Voltronics NMTM120C), which has a range of 1--125 pF, and maintains tunability even at low temperatures. A G-10 rod with a tapered end fits into the tuning shaft of the tuning capacitor. The rod extends the length of the probe, where it passes through a vacuum flange (Goddard Quikonnect valve). This allows tuning from outside of the cryostat, an important capability for low temperature experiments since the change in temperature from 300 K to 4 K causes the resonance to change appreciably.

\section{Experimental Results}
\subsection{CW DNP}
\noindent To demonstrate the performance of our probe, we ran proton CW DNP experiments on a standard sample of 20 mM TEMPO in a 60/40 glycerol-d5/D$_2$O mixture. The sample was degassed using a freeze-pump-thaw cycle, which was repeated three times. This was done to remove dissolved paramagnetic oxygen to reduce relaxation pathways and provide a nuclear relaxation time long enough to see significant DNP enhancement.
	
\begin{figure}
  \centering
    \includegraphics[width=.65\textwidth]{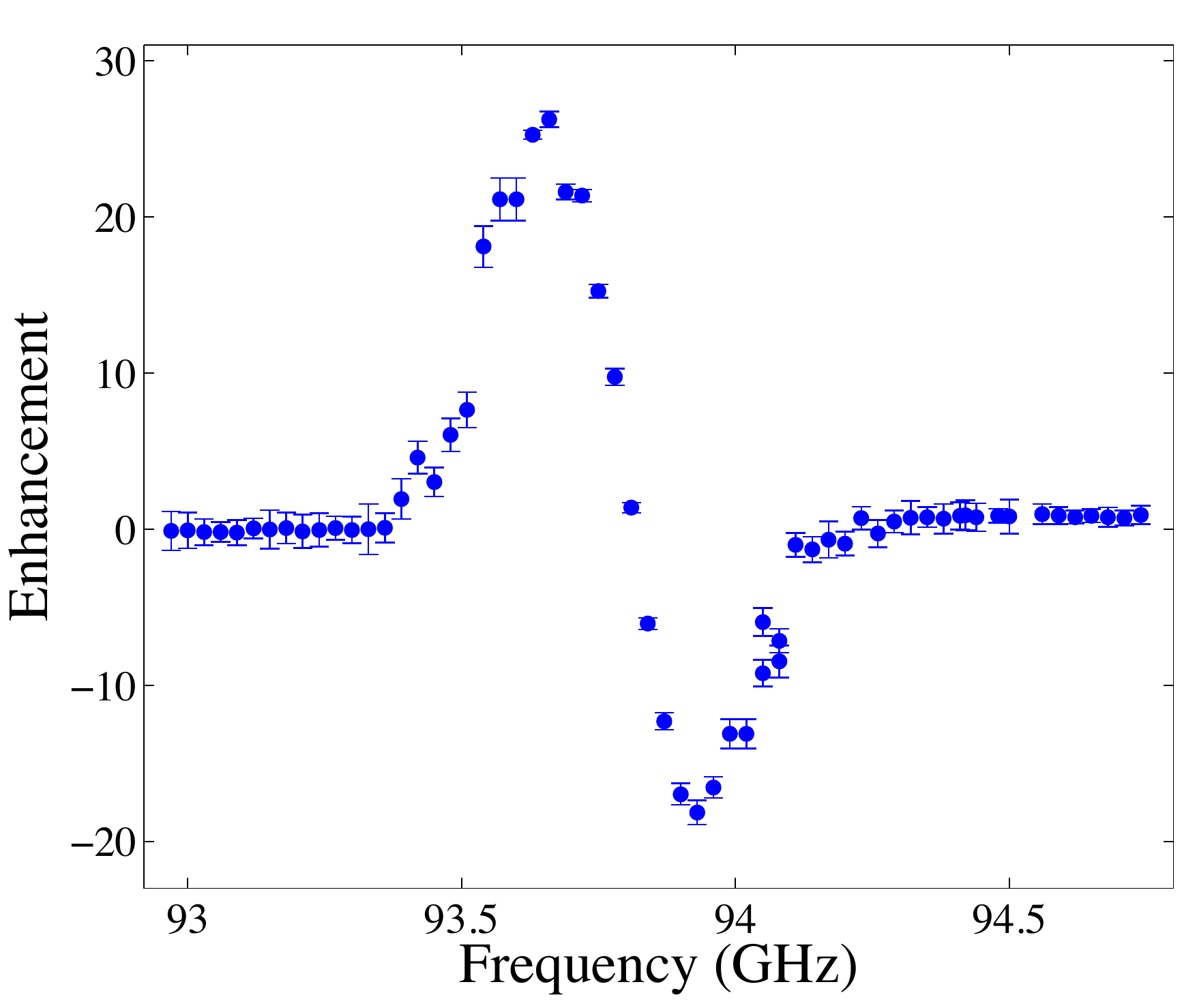}
  \caption{Proton DNP enhancement measured as a function of the microwave excitation frequency in a 20 mM solution of 4-amino TEMPO in a 60:40 glycerol-d5/D$_2$O mixture.}
  \label{fig:CWDNPexpt}
\end{figure}	

Figure~\ref{fig:CWDNPexpt} shows the DNP enhancement observed at 4.2 K as a function of microwave excitation frequency using the pyramidal horn. The reported enhancements, defined as $(\epsilon = S_{\mathrm{\mu w \: on}}/S_{\mathrm{\mu w \: off}} - 1)$, were measured following either a 60~s delay or 60~s of microwave irradiation.  In all cases a 50-pulse train was used to ensure that the spins were completely saturated at the start of the experiment.  The proton T$_1$ is about 300 s. The strength of the signal in each experiment was estimated by averaging the first 8 points of the real channel of the FID.  Errors were estimated using the standard deviation of the noise, which was taken to be points 75 to 100 in the time domain data (a signal-free region). The experiments were performed in a 3.3~T field using a Bruker DRX spectrometer. The $\pi/2$ pulse in this experiment is  approximately 1.5 $\mu s$. For this TEMPO concentration the maximal enhancement obtained was about 25.

\subsection{Frequency modulated DNP}
\begin{figure}
  \centering
    \includegraphics[width=0.8\textwidth]{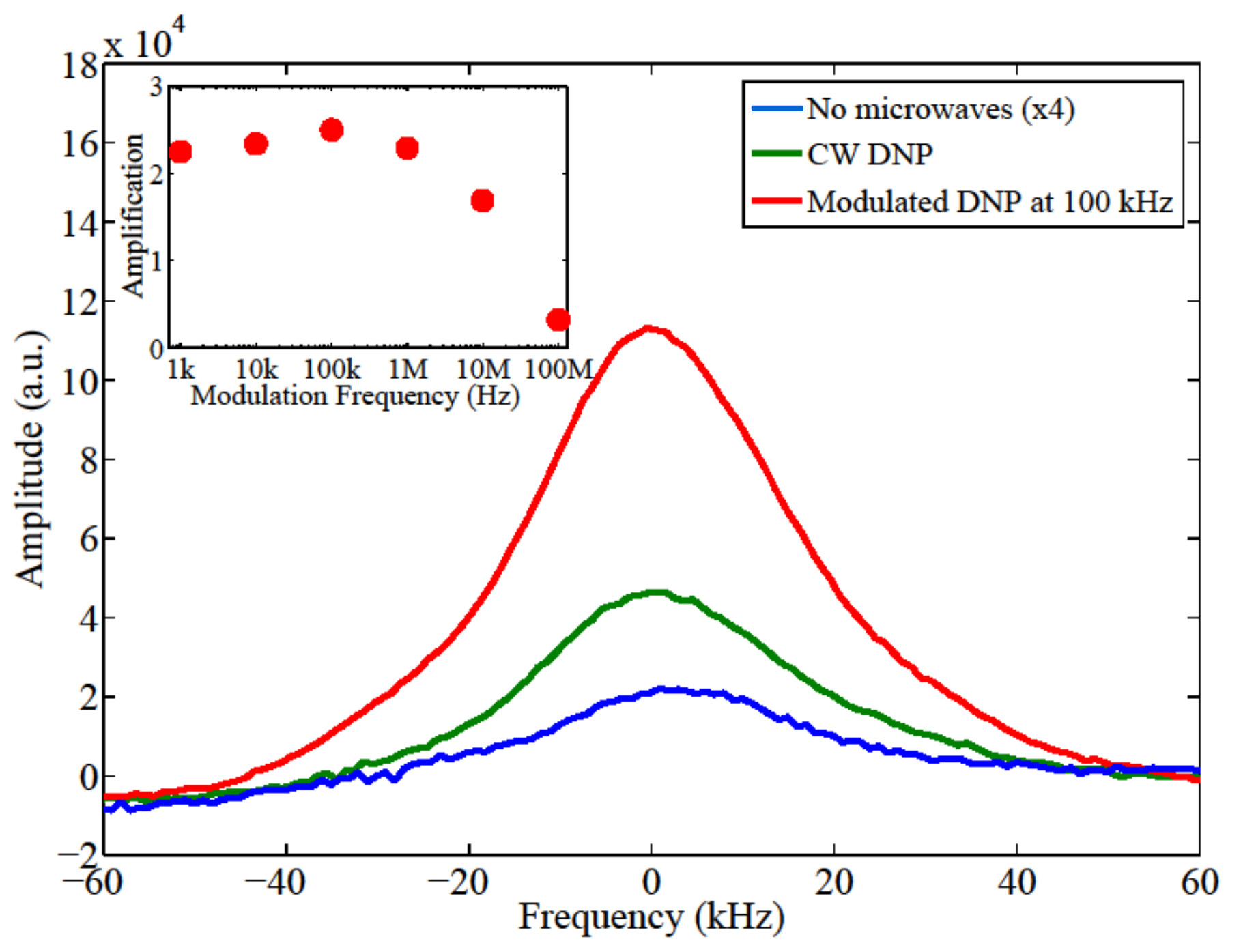}
  \caption{Comparison of the proton DNP enhancements obtained in the absence of modulation with a 100 kHz triangular frequency modulation at 4.2 K.  The center frequency for the modulation is 93.59 GHz and the microwave irradiation time is 60 s.  The modulation bandwidth is 90 MHz. The inset shows the change in amplification ($A=\epsilon_{\mathrm{mod}}/\epsilon_{\mathrm{const}}$)  with modulation frequency over 5 orders of magnitude of modulation frequency.}
      \label{fig:ModDNP}
\end{figure}

\noindent We performed a series of triangular frequency-modulated DNP experiments to demonstrate the capabilities of the system, and compared the performance against that obtained using standard constant frequency irradiation.  Figure~\ref{fig:ModDNP} shows the proton spectra obtained without DNP, with constant frequency DNP, and with frequency modulated DNP using a 100 kHz modulation rate after a 60 s build-up time.  The modulation bandwidth, which describes the range of the frequencies swept, was 90 MHz.  We define the amplification  $A = \epsilon_{\mathrm{mod}}/\epsilon_{\mathrm{const}}$ to be the ratio of the DNP enhancement (defined above) obtained using frequency modulation to that obtained using constant frequency irradiation. The inset to the figure shows the dependence of the amplification on the modulation rate (how quickly the frequency is swept) in the range from 1 kHz to 100 MHz, which is close to the proton Larmor frequency (142 MHz).  The center frequency for the modulation experiments was 93.59 GHz, which was also the frequency used for the constant frequency experiment.  Note that this frequency is offset from the frequency at which maximal CW DNP was observed in Figure~\ref{fig:CWDNPexpt}, to enhance the effects of the modulation as has previously been shown by others \cite{Hovav}.  The enhancements due to modulation depend on the concentration of the sample, the temperature, and the center frequency of the modulation.  A systematic study of these effects is beyond the scope of this paper.

\section{Conclusion}
\noindent 
We have described the design and performance of a novel millimeter wave source that has been specifically designed to engineer arbitrary waveform shaping capabilities at 94 GHz for both pulsed and CW excitation, with a bandwidth greater than 1 GHz. Such a source opens up the possibility of performing significantly more complicated modulations to enhance DNP performance.  We describe a simple overmoded waveguide transmission scheme that efficiently transmits the millimeter waves from room temperature outside the magnet to a cryogenic environment inside the magnet with a loss of about 1 dB, comparable to that achievable with corrugated waveguide and quasi-optical techniques.  
 The microwave bandwidth of the probe is limited primarily by the antenna or resonator used to couple the microwaves to the sample, while the small microwave loss observed is dominated by the quartz window used to seal the brass pipe used as a waveguide.  We demonstrated the use of this system for triangular frequency modulated DNP experiments, achieving modulation frequencies of 100 MHz, an order of magnitude higher than has previously been demonstrated.

Potential improvements include a better system of aligning the horns at the top and bottom of the cryostat to ensure the system is always in its lowest-loss configuration.  Additionally, microwave losses could be reduced even further by using a thinner and/or lower loss material for the window seal.  We are currently exploring different antenna-coil configurations that maintain the broadbandedness of the system while maximing the coupling between the fields and the sample.  Such a probe will be be invaluable for the systematic exploration of modulation schemes to improve DNP enhancement.
For example, it would be very interesting to use our broadband, high field DNP probe to try to extend some of the ideas performed at low field to higher field, such as the application of optimal control theory of DNP \cite{Sheldon}.

\section{Acknowledgements}
\noindent This material is based in part upon work supported by the National Science Foundation under CHE-1410504 and by the Pilot Project Program of the Dartmouth Physically Based Center for Medical Countermeasures Against Radiation, with NIH funding from the National Institute of Allergy and Infectious Diseases (U19-AI091173). The authors would like to thank Josef Granwehr for helpful discussions on the instrumentation, and  Whitey Adams and Chris Grant for machining various components of our set-ups and Leon from Aerowave for helpful discussions.

\section{References}

\bibliography{ProbeBib}

\end{document}